\documentclass[preprint]{revtex4}
\usepackage{graphicx}
\usepackage{subfigure}
\usepackage{multirow}
\usepackage{hyperref}
\usepackage{amsmath,bm,amssymb,amsfonts}
\usepackage{color}
\newcommand{\dalm}{\kern1pt\vbox{\hrule height 0.9pt\hbox{\vrule width 0.9pt
			\hskip 2.5pt\vbox{\vskip 5.5pt}\hskip 3pt\vrule width 0.3pt}\hrule height 0.3pt}
	\kern1pt}
\begin{document}
	\title{{\bf
			Effects of Dark Matter on the Spontaneous Scalarization in Neutron Stars}}
	
	\author{{\bf Fahimeh Rahimi $^{1,2}$ } and {\bf Zeinab Rezaei $^{1,2}$} \footnote{Corresponding author. E-mail:
				zrezaei@shirazu.ac.ir}}
	\affiliation{ $^{1}$Department of Physics, School of Science, Shiraz
University, Shiraz 71454, Iran.\\
		$^{2}$Biruni Observatory, School of Science, Shiraz
University, Shiraz 71454, Iran.}
	
	
	\begin{abstract}

Dark matter, an important portion of compact objects, can influence different phenomena in neutron stars.
The spontaneous scalarization in the scalar-tensor gravity has been proposed for neutron stars.
Here, we investigate the spontaneous scalarization in dark matter admixed neutron stars. Applying the dark matter equations of state,
we calculate the structure of scalarized neutron stars containing dark matter. The dark matter equations of state are based on observational data from
the rotational curves of galaxies and the fermionic self-interacting dark matter. Our results verify that the spontaneous scalarization is affected by the dark matter pressure in neutron stars. Depending on the central density of scalarized dark matter admixed neutron stars, the dark matter pressure alters the central scalar field. The increase of dark matter pressure in low-density scalarized stars amplifies the central scalar field. However, the pressure of dark matter in high-density scalarized stars suppresses the central scalar field. Our calculations confirm that the stars in the merger event GW170817
and in the low-mass X-ray binary 4U 1820-30 can be scalarized dark matter admixed neutron stars.

	\end{abstract}
	
	\maketitle
\section{Introduction} \label{sec:intro}

{Dark matter (DM) which can be captured by relativistic objects affects the various astrophysical phenomena
in these stars.
Scattering of the DM with nucleons and hyperons in neutron stars (NSs) concerns the transferring of large momentum
which leads to the influence of the nucleon structure, the strong interactions, and the
momentum dependence of the hadronic form factors on the DM capture rate \cite{arXiv:2012.08918,arXiv:2108.02525}.
The effective capture of DM by NSs alters the cooling process in these stars via
the DM impact on the neutrino emissivity resulting in the enhancement of the star cooling and internal relaxation rates
\cite{arXiv:2203.02132}.
Electroweak multiplet DM is effectively captured in NSs with a large elastic scattering cross-section
leading to the possibility of the temperature observation of the NSs \cite{arXiv:2204.02238}.
DM-nucleon scattering cross section enhances by the NSs, giving rise to
detecting of DM through NS spectroscopy \cite{arXiv:2104.02700}.}

{The sensitivity of NS heating to different properties of DM can be employed to explore
the DM impacts on the relativistic stars \cite{arXiv:1807.02840,arXiv:2301.08767,arXiv:1904.09803,arXiv:1906.10145,arXiv:1911.06334,arXiv:2001.09140,arXiv:2207.02221}.
Heating of NSs which takes place via the energy transfer by the inelastic DM \cite{arXiv:1807.02840,arXiv:2301.08767}
and leptophilic DM \cite{arXiv:1904.09803} may be detected using infrared telescopes.
Capturing the muonphilic DM in NSs is possible due to the existence of stable muons and degenerate electrons in these stars
leading to the NS heating up kinetically and via the annihilations \cite{arXiv:1906.10145}.
DM scattering and annihilation in the nuclear pasta of NS crust result in the heating of the NS
showing the importance of the NS crust as a thermal detector of DM \cite{arXiv:1911.06334}.
GeV Dirac fermion DM may also be captured by the NSs causing the heating of NSs via the deposited kinetic energy \cite{arXiv:2001.09140}.
Exploring the pseudoscalar-mediated DM through the DM-induced heating of NSs is more favorable
compared to the direct mediators and DM searches \cite{arXiv:2207.02221}.}

{Astrophysical observations confirm the existence of DM admixed compact stars as well as the significance of DM effects on these
objects \cite{arXiv:2002.10961,arXiv:2109.01853,arXiv:2110.05538}.
GW170817 and GW190425 merger events can be regarded as relativistic stars containing both nuclear and
bosonic self interacting DM \cite{arXiv:2002.10961}.
Relativistic mean-field models \cite{arXiv:2109.01853} and general relativistic model based on the self-interacting scalar field \cite{arXiv:2110.05538}
show that the compact component in the GW190814 merger event is a massive NS admixed
with DM particles.
Properties of DM admixed relativistic stars have been vastly investigated in the literature.
The interplay between DM and baryonic matter causes significant effects on NSs \cite{arXiv:1905.00893,arXiv:1905.12483,arXiv:2002.00594,arXiv:2212.12615}.
Energy injection from DM self-annihilations can cause quark matter droplet nucleation in DM admixed NSs (DMANSs) \cite{arXiv:1905.00893}.
Fermionic DM interacting with the nucleons via the Higgs portal grows the rate of DMANS cooling \cite{arXiv:1905.12483}.
Thermodynamic properties of symmetric nuclear matter, pure neutron matter, and NS matter are affected by DM in NSs \cite{arXiv:2002.00594}.
DM which interacts with the hadronic and quark matter via the exchange of Higgs boson in hybrid NSs changes the discontinuity on the energy density,
the star minimum mass, and the mass-radius relation of hybrid stars admixed with DM \cite{arXiv:2212.12615}.
DM also modifies the NS equation of state (EoS) \cite{arXiv:2209.10905,arXiv:2304.05100,arXiv:2306.17510,arXiv:2311.00113}.
Asymmetric bosonic DM as a core in NSs leads to the effective softening of the EoS while it causes the stiffening of the
EoS when considered as a DM halo \cite{arXiv:2209.10905}.
DM interactions in DMANSs result in the softening of the EoS and the decrease of star
maximum mass, radius, and tidal deformability \cite{arXiv:2304.05100,arXiv:2306.17510,arXiv:2311.00113}.
Moreover, the curvature of NSs and their binding energy \cite{arXiv:2007.05382}, the gravitational wave emitted from NSs
\cite{arXiv:2109.03801}, and the distribution of DM as a dense dark core or an extended dark halo in NSs \cite{arXiv:2204.05560}
are influenced by the properties of DM in these stars.
In consequence, the DMANSs can be considered as probes for constraining the DM characteristics \cite{arXiv:1910.09925,arXiv:2212.12547}.
NSs give constraints on the DM particle mass as well as the DM fractions in the star \cite{arXiv:1910.09925}.
High energy neutrino from NSs presents the bounds on the long-lived DM mediators \cite{arXiv:2212.12547}.}

{Spontaneous scalarization of NSs which is a tachyonic instability as a result of the nonminimal coupling of scalar field and curvature
can arise in the scalar-tensor gravity.
This theory of gravity predicts different properties for the NSs compared to the ones in general relativity
\cite{arXiv:1806.00568,arXiv:1808.04391,arXiv:2007.10080,arXiv:2109.13453,arXiv:2005.12758,arXiv:2010.14833,arXiv:2004.00322,arXiv:2107.07036}.
NSs in scalar-tensor gravity have maximum compactness lower than the compactness of the general relativistic stars \cite{arXiv:1806.00568}.
The x-ray pulse profile from the hotspots on the NS surface
can be influenced by the star scalar field and deviates from the one in the general relativity \cite{arXiv:1808.04391,arXiv:2007.10080,arXiv:2109.13453}.
The characteristics of NS in the scalar-tensor gravity as well as its scalarization depend on the mutual interplay between magnetic and scalar fields
of NSs \cite{arXiv:2005.12758}.
NS spontaneous scalarization alters the NS magnetic deformation and the emitted gravitational waves from these stars \cite{arXiv:2010.14833}.
Scalar tensor gravity which leads to the scalarized NSs influences the iron line from accreting NSs \cite{arXiv:2004.00322}.
For oscillating NSs, scalar-tensor gravity results in the modes in its spectrum which are different from the ones in general relativity \cite{arXiv:2107.07036}.
In addition, the observational data related to NSs are used to constrain the parameters of the scalar
tensor gravity \cite{arXiv:2105.13644,arXiv:1903.00391,arXiv:2204.02138}.
Applying the data from the NS-black hole gravitational wave events and using the Bayesian inference,
the scalar-tensor gravity can be constrained \cite{arXiv:2105.13644}.
The properties of spontaneous scalarization in scalar-tensor gravity have been constrained via
the data related to the mass and radius of NSs \cite{arXiv:1903.00391,arXiv:2204.02138}.}

{Massive scalar-tensor gravity \cite{arXiv:1707.02809,arXiv:1805.07818,arXiv:1812.00347}, scalar Gauss-Bonnet gravity \cite{arXiv:1712.03715,arXiv:2103.11999},
tensor multi-scalar theories of gravity \cite{arXiv:1911.06908,arXiv:2004.03956,arXiv:2105.08543}, multi-scalar Gauss-Bonnet gravity \cite{arXiv:2109.09399},
and degenerate higher-order scalar-tensor theories \cite{arXiv:2207.13624} have
been considered to explore the scalarized NSs.
In scalar-tensor gravity with a massive scalar field, NS maximum mass is higher than the one in general relativity \cite{arXiv:1707.02809}.
Considering the self-interacting massive scalar field, the mass of the scalar field and its self-interaction
affect the properties of scalarized NSs \cite{arXiv:1805.07818,arXiv:1812.00347}.
In extended Gauss-Bonnet scalar-tensor theories, the scalarization of NSs is due to the curvature of the spacetime
rather than the NS matter \cite{arXiv:1712.03715}.
Scalar Gauss-Bonnet gravity verifies that scalarized NSs can be formed from the core collapse of a nonscalarized star
 \cite{arXiv:2103.11999}.
Tensor multi-scalar theories of gravity forecast the spontaneous scalarization in the new classes of relativistic stars called topological
\cite{arXiv:1911.06908} and non-topological \cite{arXiv:2004.03956} NSs.
The spontaneous scalarization of non-topological NSs can also take place in multi-scalar Gauss-Bonnet gravity
 \cite{arXiv:2109.09399}.
Non-uniqueness of scalarized NSs and two solutions for scalarized stars have been reported in tensor multi-scalar theories of gravity  \cite{arXiv:2105.08543}.
In degenerate higher-order scalar-tensor theories, the mass and radius of NSs are obtained with higher values
compared to the ones in general relativity \cite{arXiv:2207.13624}.}

{Regarding the above discussions on the DM in relativistic stars and the properties of NSs in scalar-tensor gravity,
it seems that the DM can also affect the spontaneous scalarization in NSs.
In the present work, we investigate how the DM alters the NSs in scalar-tensor gravity and its spontaneous scalarization.
In Section \ref{s2}, the EoSs for the DM which are employed in this paper are introduced.
Section \ref{s3} belongs to the model description of DMANSs in the scalar-tensor gravity.
In Section \ref{s4}, we discuss the structural properties of scalarized NSs that contain the DM.
Section \ref{s5} concerns the Summary and Conclusions.}

\section{Dark Matter Equations of State}\label{s2}

\begin{figure}[h]
	\subfigure{}{\includegraphics[scale=0.85]{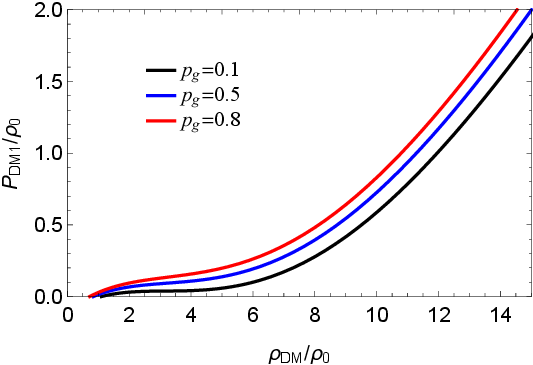}
			}
	\subfigure{}{\includegraphics[scale=0.85]{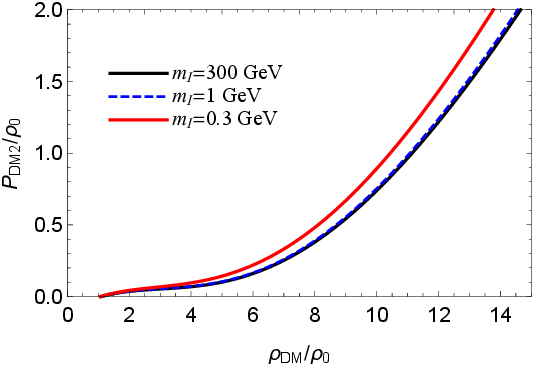}
		}	
	\caption{Left: Dark matter equation of state in the first model considering $\rho_g=0.2\times10^{16}g/cm^3$ and different values of $p_g$. In this figure and all following figures, $p_g$ is in units of $10^{35} dyn/cm^2$. Right: Dark matter equation of state in the second model with the mass $m=1 \ GeV$ and different values of the interaction between particles, $m_I$. We have assumed $\rho_0=1.66\times10^{14}g/cm^3$. }
	\label{pdm}
\end{figure}

In this paper, we explore the impacts of the DM pressure on the scalarized NSs. To express the DM
EoS, we utilize two models which describe the DM pressure. In the first model, the DM EoS is given by the observational data from the rotational curves of galaxies \cite{Barranco}. The pseudo-isothermal model results in the mass density profile which has
regularity at the origin. The velocity profile, geometric potentials, and gravitational potential give
the EoS in the pseudo-isothermal density profile with the following form \cite{Barranco},
\begin{eqnarray}\label{213}
       {P_{DM1}}({\rho_{DM}})=\frac{8  {p}_g}{\pi^2-8}[\frac{\pi^2}{8}-\frac{arctan\sqrt{\frac
       {{\rho}_g}{{\rho_{DM}}}-1}}{\sqrt{\frac{{\rho}_g}
       {{\rho_{DM}}}-1}}
	-\frac{1}{2}(arctan\sqrt{\frac
       {{\rho}_g}{{\rho_{DM}}}-1}\ )^2],
 \end{eqnarray}
with the density, $\rho_{DM}$, and the pressure, $P_{DM1}$, of DM. Besides, $\rho_g$ and $p_g$ are the free parameters denoting
the central density and pressure of galaxies. For the DM in NSs, the free parameters, $\rho_g$ and $p_g$, are of the order of the central density and the pressure of NSs, respectively \cite{Rezaei}. Figure \ref{pdm} (Left) presents the DM pressure in the first model, assuming the value
$\rho_g=0.2\times10^{16}g/ cm^{3}$ and different values of $p_g$ and also the value $\rho_0=1.66\times10^{14}g/cm^3$. The increase in $p_g$ leads to the stiffening
of the DM EoS.

In the second model, we assume the DM as an interacting Fermi gas at zero temperature containing $N$ particles with mass $m$ and spin $1/2$.
The internal energy per particle is related to the one-body term, $E_1$, and the interaction two-body term, $E_2$, as follows,
\begin{eqnarray}\label{etot}
           E_{tot}=E_1+E_2,
 \end{eqnarray}
with
\begin{eqnarray}
           E_{1}=\frac{m^4 c^5}{2 \pi^2 \hbar^3}\frac{1}{n_{DM}}\sum_{i=+,-}\frac{1}{8}\{x_F^{(i)}\sqrt{1+{x_F^{(i)}}^2}(1+2{x_F^{(i)}}^2)-sinh^{-1}(x_F^{(i)})\},
            \end{eqnarray}
in which $n_{DM}$ denotes the
total number density of DM particles and $x_F^{(i)}=\frac{\hbar k_F^{(i)}}{m c}$ with $k_F^{(i)}$ which is the Fermi momentum of a DM particle with spin projection $i$. For the interaction two-body energy, we apply the form considered in Ref. \cite{Narain},
 \begin{eqnarray}\label{e2pm}
           E_{2}=\frac{u}{n_{DM}},
            \end{eqnarray}
with the interaction energy density of the particles, $u$, supposing the spin-independent interaction one.
Considering the lowest order approximation, the interaction energy density is given by \cite{Narain},
\begin{eqnarray}
          u=\frac{{n_{DM}}^2}{m_I^2},
            \end{eqnarray}
with the energy scale of the interaction between DM
particles, $m_I$. The first law of thermodynamics,
 \begin{eqnarray}\label{pdm2}
          P_{DM2}=n_{DM}^2 (\frac{\partial E_{tot}}{\partial n_{DM}}),
            \end{eqnarray}
results in the pressure of DM in the second model. Figure \ref{pdm} (Right) shows the DM pressure of the second model
with the mass $m=1 \ GeV$ and different values of the interaction between particles, $m_I$. The DM EoS is more stiffer with the lower values of $m_I$.
Therefore, the decrease of $m_I$ which corresponds to the growth of the interaction between DM particles results in the stiffening of the EoS.

\section{Dark Matter Admixed Neutron Stars in Scalar Tensor Gravity}\label{s3}

\begin{figure}[h]
	\subfigure{}{\includegraphics[scale=0.85]{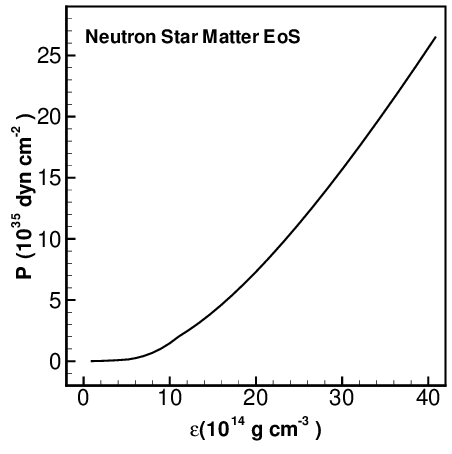}}	
	\caption{Neutron star matter equation of state constrained by the observational data \cite{Jiang}. }
	\label{nseos}
\end{figure}

Spherical symmetric static DMANS can be modeled with a spacetime line element in the scalar-tensor theory in Einstein frame,
\begin{equation}
	ds^2 = - N(r)^2 dt^2 + A(r)^2 dr^2 + r^2
	(d\theta^2 + \sin^2\theta d\phi^2),
\end{equation}
with the metric functions $N(r)$ and $A(r)= [1-2 m(r)/r ]^{-1/2}$ and the mass profile $m(r)$. Considering the action of scalar-tensor theory, the field equations lead to five differential equations \cite{28},

\begin{eqnarray}\label{}
  \frac{d m}{dr} = 4\pi r^2 a^4 \tilde{\epsilon} + \frac{r}{2} (r-2m) \Big(\frac{d\phi}{dr}\Big)^2 \label{eq:dm},
 \end{eqnarray}
\begin{eqnarray}\label{}
 \frac{d \ln N}{dr} = \frac{4\pi r^2 a^4 \tilde{p}}{r - 2m} +\frac{r}{2} \Big(\frac{d\phi}{dr}\Big)^2 + \frac{m}{r(r-2m)} \label{eq:dn},
 \end{eqnarray}
\begin{eqnarray}\label{}
\frac{d^2\phi}{dr^2} = \frac{4\pi r a^4}{r-2m} \! \left[ \alpha (\tilde{\epsilon} - 3\tilde{p}) + r (\tilde{\epsilon} - \tilde{p}) \frac{d\phi}{dr} \right ]\! -\frac{2(r-m)}{r(r-2m)} \frac{d\phi}{dr} \label{eq:dphi},
 \end{eqnarray}
\begin{eqnarray}\label{}
\frac{d\tilde{p}}{dr} = -(\tilde{\epsilon} + \tilde{p}) \left[  \frac{4\pi r^2 a^4 \tilde{p}}{r-2m} \! + \! \frac{r}{2} \Big(\frac{d\phi}{dr}\Big)^2 \!\! + \! \frac{m}{r(r-2m)} \! + \! \alpha \frac{d\phi}{dr} \right], \label{eq:dp}
 \end{eqnarray}
\begin{eqnarray}\label{}
\frac{dm_b}{dr}=\frac{4\pi r^2a^3\tilde{\rho}}{\sqrt{1-\frac{2m}{r}}}\label{eq:dm_b}.
 \end{eqnarray}
Here, $\phi$ shows the scalar field and $a(\phi)$ is the coupling function with the form $a(\phi) = e^{\frac{1}{2}\beta (\phi-\phi_0) ^2}$ in which
$\beta$ is the coupling constant, $m_b$ denotes the baryonic mass, $\alpha(\phi)=\frac{d ln a(\phi)}{d \phi}$ and $\phi_0=0$. In addition, the total energy density, $\tilde{\epsilon}$, and total
pressure, $\tilde{p}$, are presented by the energy density and pressure of visible ($V$) and dark ($D$) sectors,
   \begin{eqnarray}
     \tilde{\epsilon}(r) =\varepsilon_V(r) + \varepsilon_D(r),
 \end{eqnarray}
  \begin{eqnarray}\label{press}
   \tilde{p}(r) = p_V(r) + p_D(r).
 \end{eqnarray}
{In Eq. (\ref{press}), $p_V$ demonstrates the EoS of visible matter. In this work, to describe the visible sector, we consider a system of dense NS matter
and apply the piecewise polytropic expansion constrained by the observational data of GW170817 and the data of six low-mass X-ray binaries (LMXB) with thermonuclear bursts or the symmetry energy of the nuclear interaction \cite{Jiang}.
For this NS EoS, the polytropic form $P=K\rho^{\Gamma}$ with four pressure parameters
$\{P_1,P_2,P_3,P_4\}$ is parameterized. These four parameters are pressure at the densities of $\{1, 1.85, 3.7, 7.4\}\rho_{sat}$
with the saturation density $\rho_{sat}=2.7\times10^{14}\ g\ cm^{-3}$ \citep{Ozelprd}. Joint analysis results in the constraints on the pressure parameters.
Nuclear constraints, gravitational wave data and the LMXB sources with thermonuclear bursts, LMXB source data
and the current bounds of $M_{TOV}$, and LMXB sources with thermonuclear burst mainly determine the values of $P_1$, $P_2$, $P_3$ , and
$P_4$, respectively. These values are $P_1=3.9\times10^{33}$, $P_2=1.4\times10^{34}$, $P_3=2.0\times10^{35}$ , and
$P_4=1.4\times10^{36}$ in units of $dyn\ cm^{-2}$. Figure \ref{nseos} shows this EoS of NS matter.}

Besides, $p_D$ in Eq. (\ref{press}), denotes the DM pressure which is chosen from Eq. (\ref{213}) and Eq. (\ref{pdm2}), in the first and second model, respectively.
We solve the equations employing the boundary conditions,
\begin{eqnarray}\label{}
&m(0) =m_b(0)= 0, \quad \lim_{r\to\infty}N(r) = 1,\quad \phi(0)=\phi_c, \quad \lim_{r\to\infty}\phi(r) = 0, \nonumber \\
	&\frac{d\phi}{dr}(0) = 0, \qquad \tilde{p}(0) = p_c, \qquad \tilde{p}(R_s) = 0. \label{eq:bc}
 \end{eqnarray}
Here, $R_s$ shows the radius of the star and $c$ denotes the center of the star.
Starting with the appropriate boundary condition ($\phi(0)=\phi_c$) at the center of star
and the iteration on $\phi_c$ with the condition \cite{28,26},
\begin{equation} \label{eq:constraint on central of scalar field}
	\phi_s  + \frac{2 \psi_s}{\sqrt{\dot{\nu}_s^2+4\psi_s^2}} \textrm{arctanh} \left[ \frac{\sqrt{\dot{\nu}_s^2 +4\psi_s^2}}{\dot{\nu}_s +2/R_s} \right] = 0,
\end{equation}
we solve the differential equations. In the above equation, s presents the surface of the star and $\psi_s = (d\phi/dr)_s$ together with $\dot{\nu}_s = 2(d\ln N/dr)|_s = R_s \psi_s^2 + 2 m_s/[R_s(R_s-2m_s)]$, and the ADM mass, $M_{ADM}$,
\begin{eqnarray}\label{}
	M_{ADM} &= \frac{R_s^2 \dot{\nu}_s}{2} \left( 1-\frac{2m_s}{R_s} \right)^\frac{1}{2}
	\exp \left[ \frac{-\dot{\nu}_s}{\sqrt{\dot{\nu}_s^2+4\psi_s^2}} \textrm{arctanh} \left( \frac{\sqrt{\dot{\nu}_s^2+4\psi_s^2}}{\dot{\nu}_s +2/R_s} \right) \right],
 \end{eqnarray}
and the scalar charge, $\omega$,
\begin{eqnarray}\label{}
\omega & = - 2 M_{ADM} \psi_s/\dot{\nu}_s.
 \end{eqnarray}
In this paper, the ADM mass of DMANSs in the scalar-tensor gravity is reported.

 \section{Results and Discussion}\label{s4}
\subsection{Mass versus the central density of dark matter admixed neutron star}

\begin{figure}[h]
\subfigure{}{\includegraphics[scale=0.85]{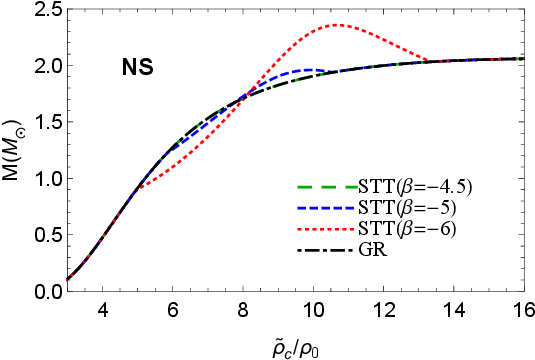}
		}
	\subfigure{}{\includegraphics[scale=0.85]{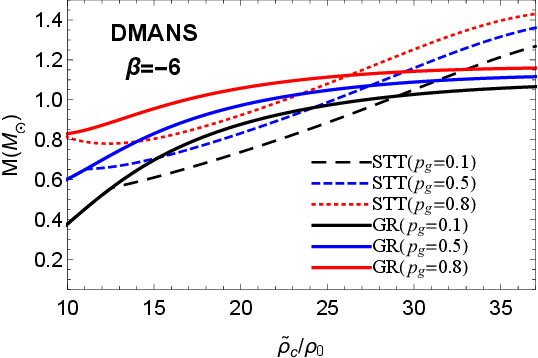}
			}
	\subfigure{}{\includegraphics[scale=0.85]{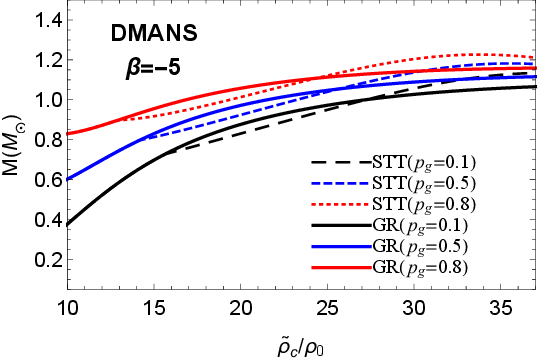}
		}	
\subfigure{}{\includegraphics[scale=0.85]{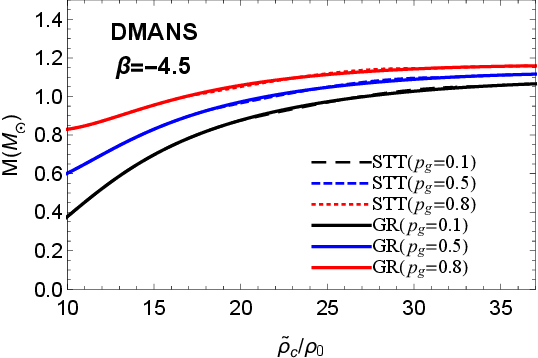}
		}	
	\caption{Mass as a function of the central density, $\tilde{\rho}_{c}$, for neutron star (NS) and dark matter admixed neutron star (DMANS) in the first model of DM EoS considering the scalar-tensor theory (STT) with different values of the coupling constant, $\beta$. The results of the general theory of relativity (GR) are also presented.}
\label{mro1}
\end{figure}

\begin{figure}[h]
\subfigure{}{\includegraphics[scale=0.85]{M-rho-NSM.eps}
		}
	\subfigure{}{\includegraphics[scale=0.85]{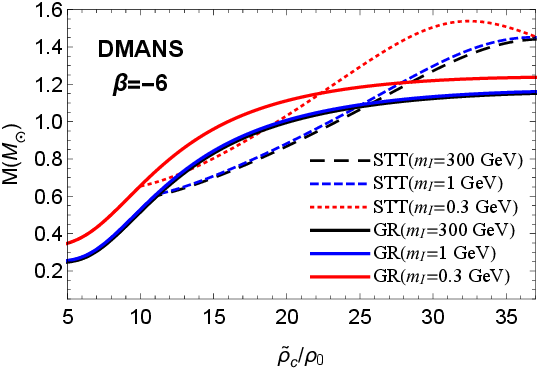}
			}
	\subfigure{}{\includegraphics[scale=0.85]{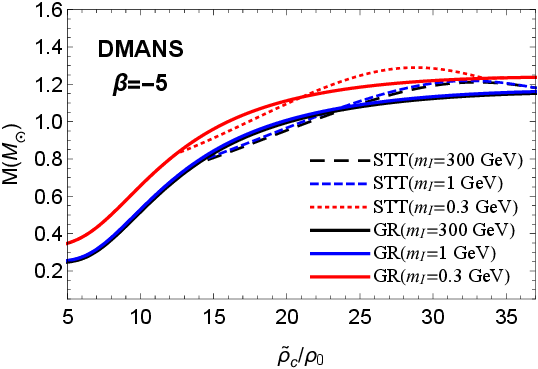}
		}	
\subfigure{}{\includegraphics[scale=0.85]{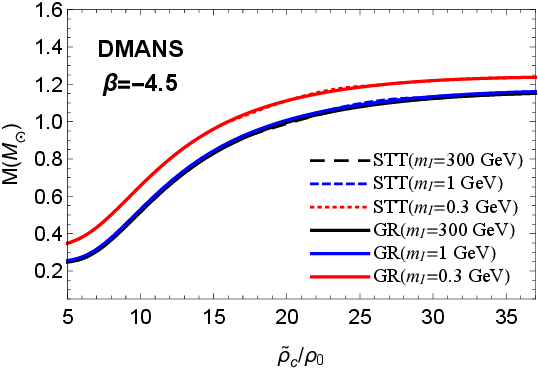}
		}
	\caption{Same as Figure \ref{mro1} but for the second model of DM EoS.}
\label{mro2}
\end{figure}

{Our results for the star mass at different central densities in the cases of NSs and DMANSs
have been given in Figures \ref{mro1} and \ref{mro2}. For NSs with the value of $\beta=-4.5$, the results of the general theory of relativity (GR)
and scalar-tensor theory (STT) are the same. Therefore, the stars with no DM and higher values of $\beta$ can not be scalarized and no scalarization takes place.
This is while for the DMANSs in both DM EoS models, the results of STT deviate from the GR ones even with $\beta=-4.5$ and these stars can be scalarized.
Hence, DM facilitates the star scalarization allowing the existence of scalarized stars with high values of the coupling constant.
Considering the lower values of $\beta$, i.e. $\beta=-5$ or $\beta=-6$, the deviation of STT branches from GR ones in DMANSs starts from the higher values of central densities compared to NSs. Moreover, the DMANSs can be scalarized even up to high densities, unlike the NSs.
Figures \ref{mro1} and \ref{mro2} show that the maximum mass of DMANSs is smaller than the NS one, due to the softening of the EoS by DM.
This is in agreement with the results reported in \cite{arXiv:2304.05100,arXiv:2306.17510,arXiv:2311.00113}. }

We find from Figure \ref{mro1} that in the first model for DM pressure, the mass is higher with stiffer DM EoS (larger $p_g$), for both
STT and GR.
{It should be noted that the lower values of $p_g$, e.g. $0.1\times10^{35} dyn/cm^2$, which correspond to softer DM EoSs lead to more reduction of the star mass.
Thus the DMANSs with smaller $p_g$ are more different from NSs in mass compared to stiffer DM EoSs.}
Considering the lower values of the coupling constant, the deviation of STT from GR in DMANS is more significant.
The stars that have different behaviors in STT and GR are scalarized DMANSs.
The stiffening of the DM EoS affects this deviation. In fact, with larger $p_g$ and stiffer DM EoS, the STT and GR branches separate at lower densities. The amount of STT deviation increases by growing $p_g$.
This means that the higher pressure of DM amplifies the star scalarization. The rate of mass growth versus the density is larger when the stars are scalarized.
Figure \ref{mro1} also verifies that most massive stars are the ones in STT with lower values of $\beta$ and stiffer DM EoS.

Figure \ref{mro2} confirms that in the second model, the decrease in  $m_I$ and the DM with higher interaction between particles give rise to more massive stars.
The increase of mass through the reduction from $m_I=1\ GeV$ to $m_I=0.3\ GeV$ is more remarkable compared to the one from $m_I=300\ GeV$ to $m_I=1\ GeV$.
The effects of DM pressure on the star mass are notable when the stars are scalarized.
The most massive stars are the scalarized ones with lower coupling constant and stronger interaction between DM particles.
DMANSs with higher interaction between particles become scalarized at lower densities.
In the case of smaller $m_I$ and higher interaction in which the EoS is stiffer, the scalarization of the star is larger.
The mass of scalarized stars grows with density more rapidly compared to GR ones. In scalarized stars, the interaction between DM particles
enlarges the mass growth with the density.

\subsection{Mass-radius relation of dark matter admixed neutron star}

\begin{figure}[h]
\subfigure{}{\includegraphics[scale=0.85]{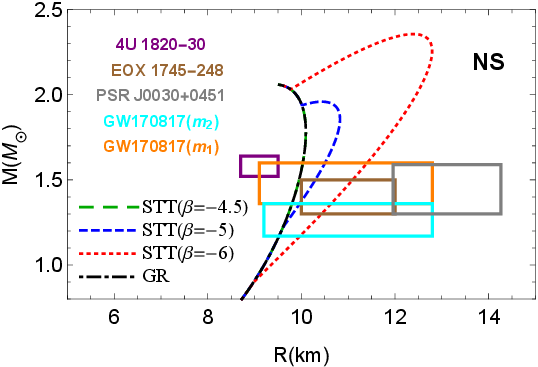}
		}
	\subfigure{}{\includegraphics[scale=0.85]{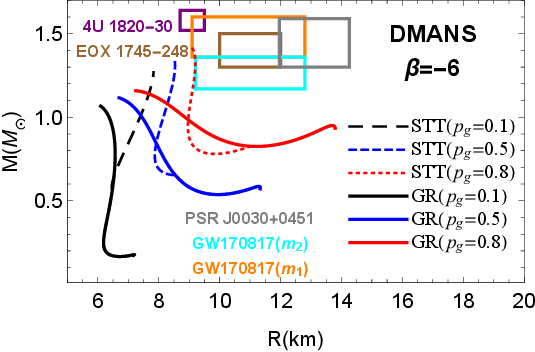}
			}
	\subfigure{}{\includegraphics[scale=0.85]{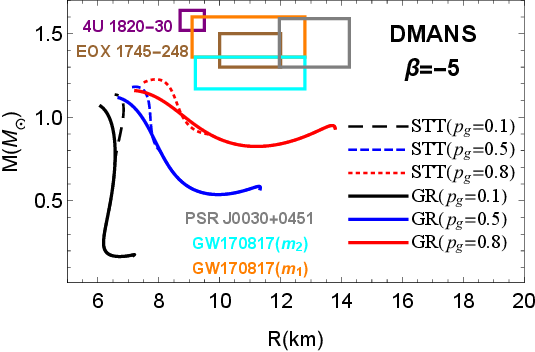}
		}	
\subfigure{}{\includegraphics[scale=0.85]{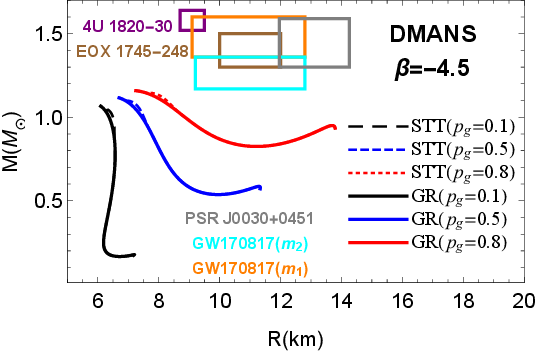}
		}
\caption{Mass versus the radius for NS and DMANS in the first model of DM EoS in GR and STT with different values of the coupling constant, $\beta$. Observational constraints on the mass and radius of NS are also given. The constraints are related to EXO 1745-248, 4U 1820-30, GW170817, and PSR J0030+0451. For more details, see the text.}
\label{mr1}
\end{figure}

\begin{figure}[h]
\subfigure{}{\includegraphics[scale=0.85]{M-R-NSM.eps}
		}
	\subfigure{}{\includegraphics[scale=0.85]{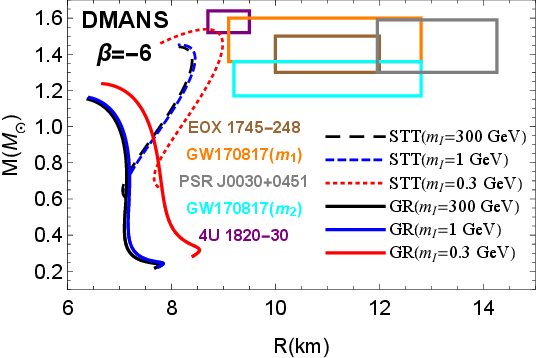}
			}
	\subfigure{}{\includegraphics[scale=0.85]{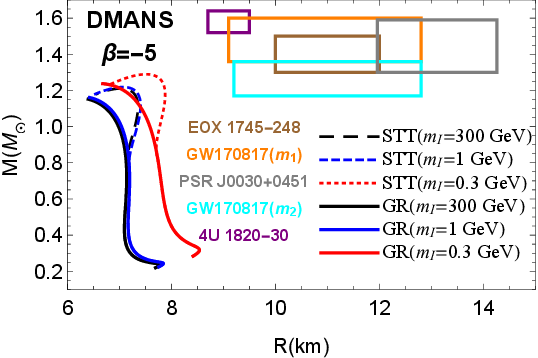}
		}	
\subfigure{}{\includegraphics[scale=0.85]{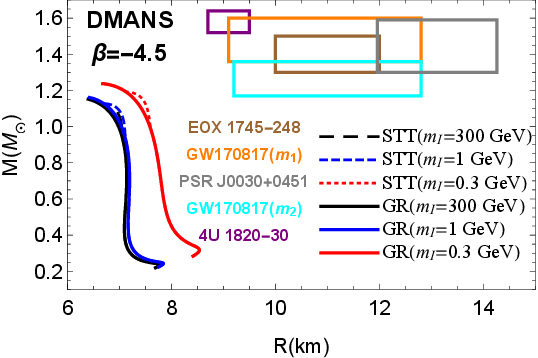}
		}	
\caption{Same as Figure \ref{mr1} but for the second model of DM EoS.}
\label{mr2}
\end{figure}

{In Figures \ref{mr1} and \ref{mr2}, we have shown the mass versus the radius for the NSs and DMANSs in STT and GR.
The observational constraints on the NS mass and radius related to EXO 1745-248 \cite{Ozel}, 4U 1820-30 \cite{Guver}, GW170817 \cite{Abbott7,Abbott8}, and PSR J0030+0451 \cite{Miller} are also presented.
The mass-radius relation of NSs in STT with $\beta=-4.5$ coincides with the result of GR, as expected because of the absence of scalarization in NSs considering the higher values of $\beta$.}
{However, for the massive scalarized DMANSs with $\beta=-4.5$, the mass-radius relation deviates from GR one.
Thus, the DM affects this relation in STT and GR, differently.
The mass-radius relation of NSs in GR and STT satisfies the constraints from EXO 1745-248 and GW170817.
The scalarization of NSs leads to the larger masses and radii of stars.}

Figure \ref{mr1} confirms that in the first model, the level of DM EoS stiffening alters the mass-radius relation for both GR and STT cases. The increase in DM pressure causes a larger radius for both nonscalarized and scalarized stars. For all DMANSs in both gravities, the range of star radius is more extended considering stiffer DM EoS, i.e. larger $p_g$. Besides, high DM pressure predicts the existence of stars with the same mass but with two sizes. It is clear from Figure \ref{mr1} that the minimum mass of stars grows by increasing the pressure of DM.
The deviation of scalarized stars from nonscalarized ones in the mass-radius relation depends on the value of $p_g$.
In DMANSs with smaller DM pressure, the scalarization results in a larger radius for massive stars, \textbf{similar to NSs}. This is while with higher DM pressure, for the stars with lower masses, the scalarized stars are smaller than GR ones, and for massive stars, the scalarization leads to larger sizes.
With lower values of the coupling constant, the scalarized DMANSs can be more massive and larger. The case of $\beta=-6$ along with $p_g=0.8\times10^{35} dyn/cm^2$ leads to scalarized stars which satisfy the constraints related to the merger event GW170817.
{Comparison of NSs with DMANSs in the first model of DM EoS approves that the DM extends the range of star radius allowing both smaller and larger DMANSs compared to NSs.}

Figure \ref{mr2} verifies that the interaction between DM particles raises the radius of stars in STT as well as GR. In most stars, the scalarization of stars also makes the stars larger. Scalarized DMANSs with lower values of the coupling constant and higher interactions between DM particles can have larger masses and radii. Considering higher values of $\beta$, the branches of nonscalarized and scalarized stars match for massive stars. However, for $\beta=-6$, the massive stars are scalarized and the results of STT deviate from GR ones. With lower values of the coupling constant and higher interactions between DM particles, two scalarized stars can exist with the same radius but different masses. Our calculations verify that the scalarized stars with $\beta=-6$ and high interactions between DM particles, i.e. $m_I=0.3\ GeV$, are in agreement with the neutron star in the low-mass X-ray binary 4U 1820-30. Accordingly, the scalarized DMANSs in two models of DM EoS are consistent with the given observational data on the assumption that the coupling constant is lower and the DM pressure is higher.
{Figure \ref{mr2} also confirms that the DMANSs in the second model of DM EoS are smaller in size compared to NSs.}

 \subsection{Central scalar field of dark matter admixed neutron star}

\begin{figure}[h]
\subfigure{}{\includegraphics[scale=0.85]{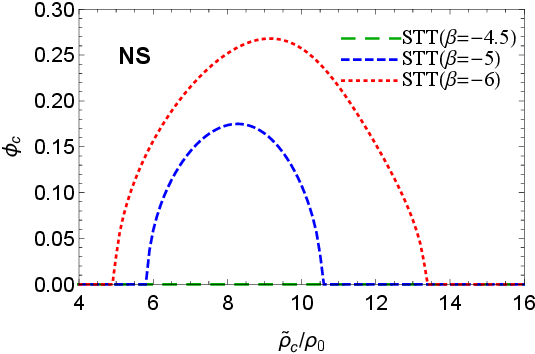}
		}
	\subfigure{}{\includegraphics[scale=0.85]{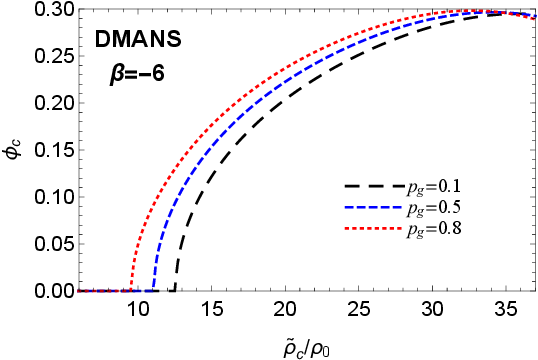}
			}
	\subfigure{}{\includegraphics[scale=0.85]{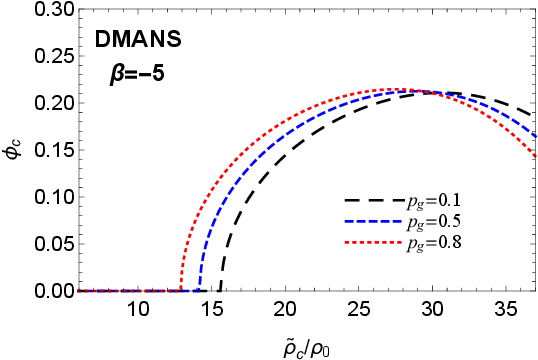}
		}	
\subfigure{}{\includegraphics[scale=0.85]{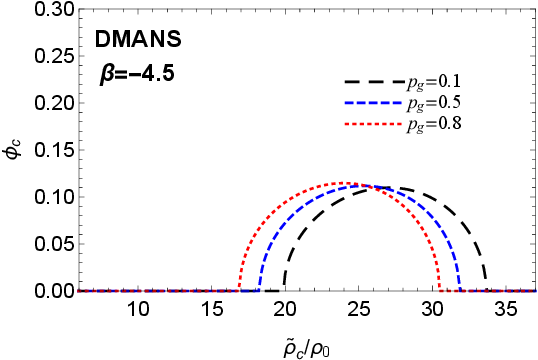}
		}	
\caption{The values of the scalar field at the center of the star, $\phi_c$, versus the central density, $\tilde{\rho}_{c}$, for
NS and DMANS in the first model of DM EoS considering different values of the coupling
constant, $\beta$.}
\label{phi1}
\end{figure}

\begin{figure}[h]
\subfigure{}{\includegraphics[scale=0.85]{phic-NSM.eps}
		}
	\subfigure{}{\includegraphics[scale=0.85]{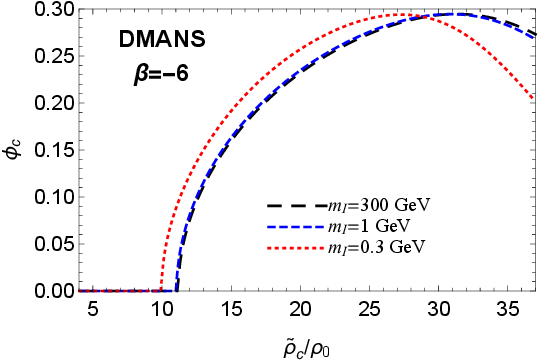}
			}
	\subfigure{}{\includegraphics[scale=0.85]{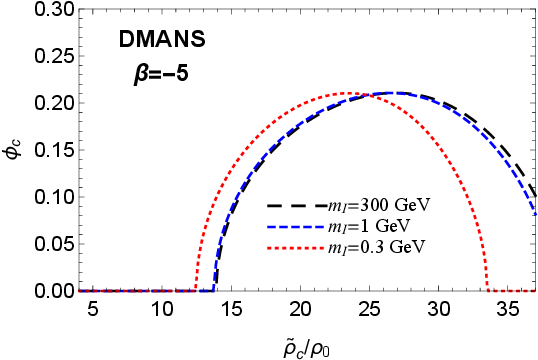}
		}	
\subfigure{}{\includegraphics[scale=0.85]{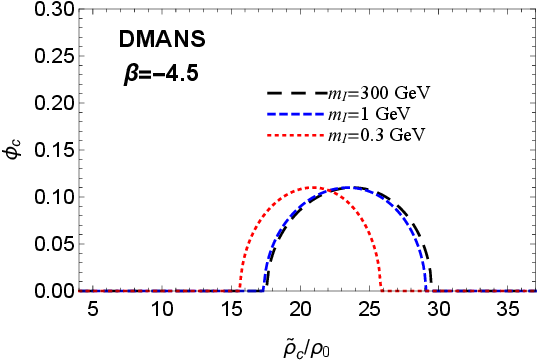}
		}	
\caption{Same as Figure \ref{phi1} but for the second model of DM EoS.}
\label{phi2}
\end{figure}

{In this part, we study the scalarization of NSs and DMANSs by exploring the scalar field at the center of the star.
Figures \ref{phi1} and \ref{phi2} present the variation of the central scalar field versus the central density for NSs and DMANSs in two models of DM EoS.}
{For NSs in STT with $\beta=-4.5$, the central scalar field equals to zero at different densities and no scalarization appears. This is while $\phi_c$ of DMANSs with $\beta=-4.5$ can be nonzero and scalarization takes place. It means that the DM induces the scalar field and scalarization in stars even with higher values of the coupling constant.}
{For the stars, at a special value of the density, the scalarization begins
and the scalar field rises from zero.}
{This critical density decreases by the reduction of $\beta$ for both NSs and DMANSs.
In the cases of $\beta=-5$ and $\beta=-6$, the NS scalar field in high-density stars becomes zero, unlike the DMANSs that are scalarized and experience the nonzero scalar field even up to high densities.}
{We can find from Figures \ref{phi1} and \ref{phi2} that the maximum value of the central scalar field for NSs at each coupling constant is smaller than the one in DMANSs. Therefore, the DM intensifies the scalar field in the center of DMANSs.}

Figure \ref{phi1} verifies that in the first model of DM EoS, the central scalar field increases by growing the density so that
with the coupling constant $\beta=-6$, the stars with the highest central densities are scalarized. Considering the stars with larger coupling constant, the central scalar field reaches a maximum value and afterward, it reduces by increasing the density. At the lower values of the density in the scalarized stars,
the central scalar field grows as $p_g$ increases. Therefore, the DM with higher pressure intensifies the central scalar field. This is while with a large enough coupling constant and at high density in the scalarized stars, the scalar field drops by $p_g$. This means that in high-density scalarized stars, the stiffer the DM EoS is, the smaller the value of the central scalar field becomes. In the case of $\beta=-6$, the effects of the DM pressure are more significant at lower values of the density in the scalarized stars.
{The values of the critical density, $\tilde{\rho}_{crit}$, for NS and DMANS are given in Tables \ref{table1} and \ref{table2}.
The critical density of scalarization in NSs is lower than the one related to DMANSs. It means that in the presence of DM, the star should be more dense to be scalarized.}
The critical density decreases by $p_g$. The higher pressure of DM results in the easier enhancement of the scalar field and the star scalarization.
Table \ref{table1} also shows that the coupling constant also alters the critical density so that it grows by increasing $\beta$.
Figure \ref{phi1} shows that the maximum value of $\phi_c$ gets larger by increasing the DM pressure. Besides, the density corresponding to the maximum scalar field reduces with $p_g$. In DMANSs with $\beta=-4.5$, at some densities, the scalar field becomes zero and the scalarization terminates. With higher values of the DM pressure, the density corresponding to zero scalar fields is smaller. Consequently, the DM pressure destroys the star scalarization.

Figure \ref{phi2} confirms that in the second model of DM EoS, the central scalar field rises from zero and gets to a maximum. Afterwards, $\phi_c$ reduces by increasing the density. At the lower values of $\beta$, the scalar field remains nonzero to high densities, and the stars with high central densities are scalarized.
However, with the coupling constant $\beta=-4.5$, $\phi_c$ vanishes at high densities. DM EoS also alters the central scalar field. In scalarized DMANSs with low central densities, $\phi_c$ increases by the reduction of $m_I$. Therefore, the interaction between DM particles amplifies the scalar field in these stars. Nevertheless, in high-density scalarized stars, the stiffer DM EoS results in a decrease in the scalar field. This behavior of DM pressure in the second model of DM EoS at low and high densities is similar to the one in the first model. Considering the DMANSs in the second model of DM EoS, the critical density decreases by the DM pressure. The higher interaction between DM particles gives rise to a nonzero central scalar field and the scalarization at lower densities. Table \ref{table2} presents the critical densities in the second model of DM EoS. The decrease of $m_I$ from $1\ GeV$ to $0.3\ GeV$ has more significant effects on the critical density. The density corresponding to the maximum central scalar field decreases by $m_I$ reduction. Noting the scalarized DMANSs with $\beta=-4.5$, the scalar field reaches zero at a special density and the stars become nonscalarized once more. The density at which the scalarization disappears decreases by increasing the DM pressure. This reduction means that the DM pressure
and the interaction between DM particles assist in the suppression of scalarization. Our results emphasize that the range of the scalarization shrinks by decreasing the value of $m_I$. From Figure \ref{phi2} we find that the maximum value of the scalar field is not importantly affected by $m_I$.
For two models of DM pressure, the stars are scalarized in larger ranges of the central density when the coupling constant is lower.

\begin{table}[h!]
	\begin{center}
		\begin{tabular}{|c@{\hspace{3mm}}|c@{\hspace{3mm}}|c@{\hspace{3mm}}c@{\hspace{3mm}}c@{\hspace{3mm}}|}
			\hline
			 & NS & &DMANS&\\	
			\hline
& & &$p_g (10^{35}\ dyn/cm^{2})$&\\	
			$\beta$ & & $0.1$&$0.5$&$0.8$\\	
			\hline
{$-6$} &5.0 &12.6 &11.1 &9.6   \\
{$-5$} &5.9 &15.7 &14.2 & 13.0  \\
{$-4.5$} & No scalarization &19.9 &18.3 &16.9   \\
					\hline\hline
				\end{tabular}
			\end{center}
	\caption{Critical density, $\tilde{\rho}_{crit}/\rho_0$, at which the scalarization takes place for
NS and DMANS in the first model of DM EoS with different values of the coupling
constant, $\beta$.}
	\label{table1}
\end{table}

\begin{table}[h!]
	\begin{center}
		\begin{tabular}{|c@{\hspace{3mm}}|c@{\hspace{3mm}}|c@{\hspace{3mm}}c@{\hspace{3mm}}c@{\hspace{3mm}}|}
			\hline
			& NS  & &DMANS&\\	
			\hline
&  & &$m_I (GeV)$&\\
			$\beta$&  & $300$&$1$&$0.3$\\	
			\hline
{$-6$}&5.0 &11.2  &11.1 & 10.1  \\
{$-5$}& 5.9&14.0  & 13.8&12.5   \\
{$-4.5$}& No scalarization &17.6  &17.4 &15.7   \\
					\hline\hline
				\end{tabular}
			\end{center}
	\caption{Same as Table \ref{table1} but for the second model of DM EoS.}
	\label{table2}
\end{table}

\subsection{Scalar charge in dark matter admixed neutron star}

\begin{figure}[h]
\subfigure{}{\includegraphics[scale=0.85]{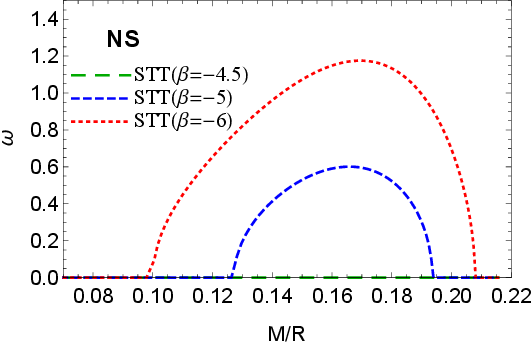}
		}
	\subfigure{}{\includegraphics[scale=0.85]{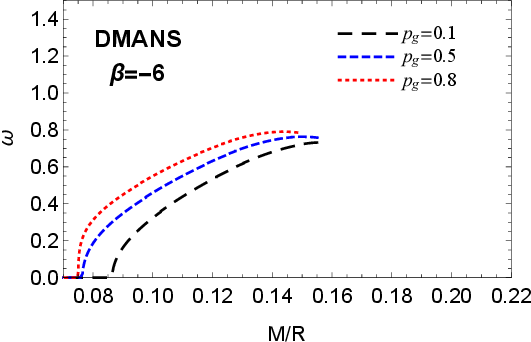}
			}
	\subfigure{}{\includegraphics[scale=0.85]{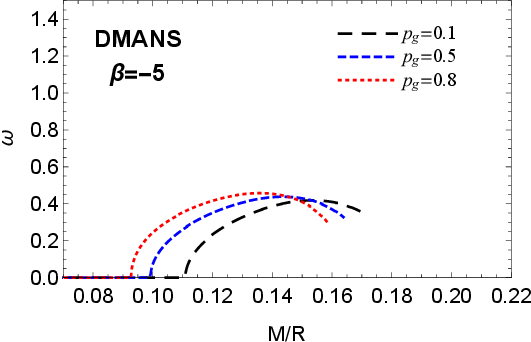}
		}	
\subfigure{}{\includegraphics[scale=0.85]{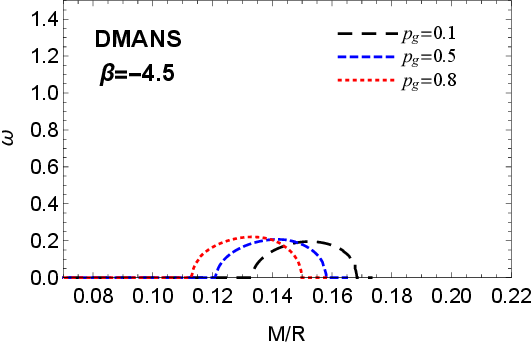}
		}	
\caption{Scalar charge, $\omega$, as a function of the compactness, M/R,
for NS and DMANS in the first model of DM EoS considering different values of the coupling
constant, $\beta$.}
\label{charge1}
\end{figure}

\begin{figure}[h]
\subfigure{}{\includegraphics[scale=0.85]{Compact-NSM.eps}
		}
	\subfigure{}{\includegraphics[scale=0.85]{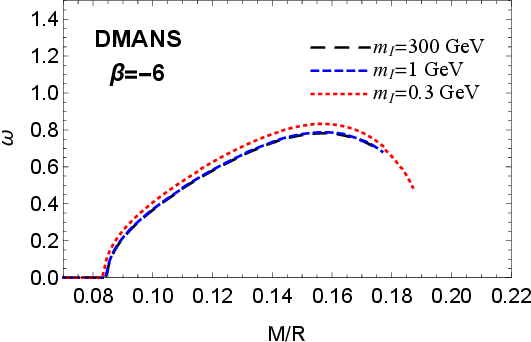}
			}
	\subfigure{}{\includegraphics[scale=0.85]{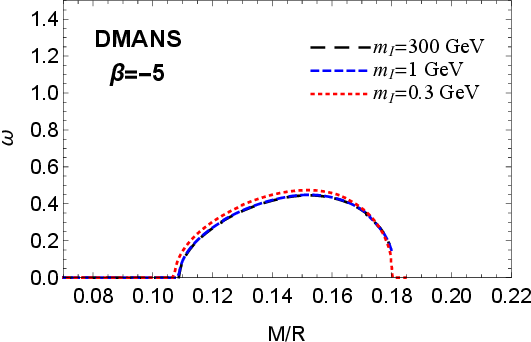}
		}	
\subfigure{}{\includegraphics[scale=0.85]{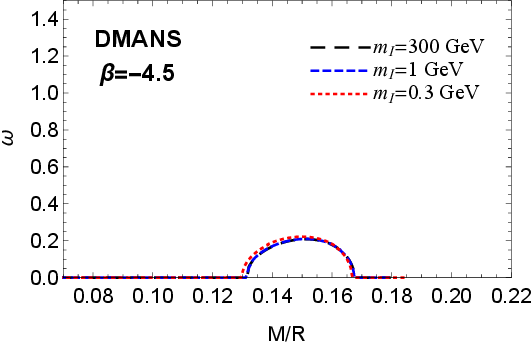}
		}	
\caption{Same as Figure \ref{charge1} but for the second model of DM EoS.}
\label{charge2}
\end{figure}

{In Figures \ref{charge1} and \ref{charge2}, we have plotted the scalar charge, $\omega$, of NSs and DMANSs applying two models of DM EoS.
Considering the case of $\beta=-4.5$, the scalar charge of NSs is equal to zero, while the DM in DMANSs results in the appearance of a star scalar charge.
With the values of $\beta=-5$ and $\beta=-6$ for the coupling constant, the star compactness which is needed to grow the scalar charge from zero is lower in DMANSs. Hence, the DM assists in the increase of the scalar charge even in the stars with lower compactness.}
{However, the maximum scalar charge of DMANSs is lower than the one of the NSs.}
We find from Figure \ref{charge1} that in the first model of DM, the scalar charge grows versus the star compactness.
With a higher coupling constant, the scalar charge can decrease and vanish as the compactness rises. At the lower values of the compactness, the scalar charge is larger in stars with higher values of $p_g$ and stiffer DM EoSs. For more compact stars, $\omega$ reduces as DM pressure increases.
With a special value of $M/R$, the scalar charge gets a maximum value which this value of $\omega$ increases by $p_g$. The corresponding compactness to the maximum scalar charge reduces with the DM pressure. In stars with larger $p_g$ and lower values of the coupling constant, the highest values of the scalar charge can be reached. In each case, at a critical value of the compactness, the scalar charge grows from zero. The critical compactness decreases by $p_g$. This indicates that with higher DM pressure, even the stars with lower compactness can experience the scalar charge.

The scalar charge of DMANSs in the second model of DM EoS is presented in Figure \ref{charge2}. Almost in all stars, the scalar charge increases by decreasing $m_I$. The interaction between DM particles leads to a larger scalar charge in DMANSs. The critical compactness at which the scalar charge grows from zero has smaller values when the interaction between DM particles and therefore the DM pressure increases. In two models of DM EoS, the critical compactness of the star grows by increasing the coupling constant. In addition, the range of the star compactness that results in the nonzero scalar charge of the stars is more extended at lower values of $\beta$.

\subsection{Critical density of scalarization in dark matter admixed neutron star}

\begin{figure}[h]
	\subfigure{}{\includegraphics[scale=0.85]{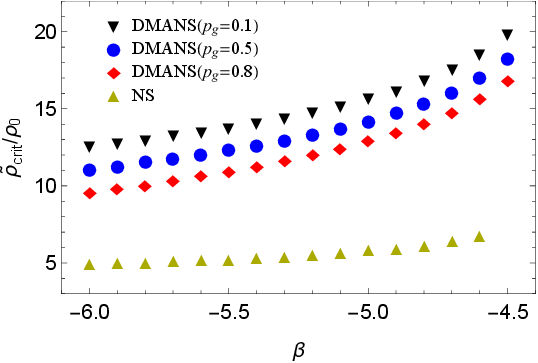}
			}
	\subfigure{}{\includegraphics[scale=0.85]{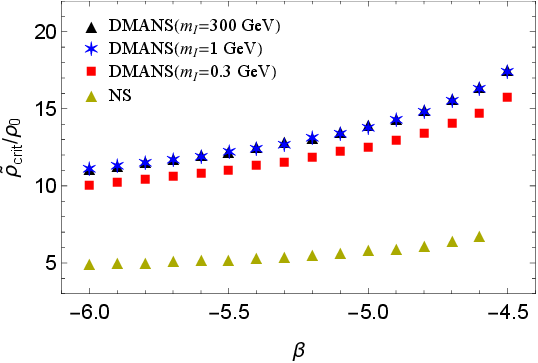}
		}	
\caption{The critical density of scalarization, $\tilde{\rho}_{crit}$, versus the coupling constant, $\beta$,
for NS and DMANS in the first (Left) and second (Right) model of dark matter pressure. The parameters of DM models are the same as Figure \ref{pdm}.}
\label{critro}
\end{figure}

In the last part of this paper, we sum up the results related to the critical density of scalarization, $\tilde{\rho}_{crit}$. Figure \ref{critro}
belongs to the critical density of scalarization \textbf{of the NSs and DMANSs} in two models of the DM pressure.
{Considering both NSs and DMANSs with DM EoSs in two models, $\tilde{\rho}_{crit}$ increases as the coupling constant grows. Hence, the stars with larger values of $\beta$ ought to have more central densities to be scalarized.}
{The critical density in NSs increases by the coupling constant very smoothly.}
In the first model of DM EoS, the critical density reduces by increasing $p_g$ and the DM pressure as discussed above and presented in Figure \ref{phi1} and Table \ref{table1}. Considering the first model, the influence of the DM EoS on the critical density is more important at lower values of $\beta$. Moreover, the critical density of the second model of DM EoS becomes smaller as $m_I$ decreases and the interaction between DM particles grows, see also Figure \ref{phi2} and Table \ref{table2}. However, in the second model, the pressure of DM alters the critical density more significantly when the coupling constant is larger.

\section{SUMMARY AND CONCLUDING REMARKS}\label{s5}

In this paper, we have studied the scalarization of dark matter admixed neutron stars in scalar-tensor gravity. The dark matter equation of state based on the observational data from the rotational curves of galaxies (first model) and the equation of state of the fermionic self-interacting dark matter (second model) have been applied to describe the dark sector in neutron stars. With higher-pressure dark matter, the scalarization of neutron stars is more notable.
The influence of dark matter pressure on the central scalar field depends on the central density.
In low-density neutron stars, the central scalar field increases by the dark matter pressure, while it decreases by the dark matter pressure in high-density scalarized neutron stars. The scalarized stars containing dark matter with higher pressure are the most massive ones.
In the first model, the effects of scalarization on the radius of neutron stars depend on the stiffening of the dark matter equation of state.
However, in the second model, the scalarized stars are larger than the nonscalarized ones.
Our results in the first model show that the stars in the merger event GW170817 can be scalarized dark matter admixed neutron stars.
Besides, the scalarized stars in the second model satisfy the constraints of the neutron star in the low-mass X-ray binary 4U 1820-30.


\acknowledgements{The authors wish to thank the Shiraz University Research Council.}


\begin{thebibliography}{}

\bibitem{arXiv:2012.08918} N. F. Bell, G. Busoni, T. F. Motta, et al., Phys. Rev. Lett. \textbf{127}, 111803 (2021);  Phys. Rev. Lett. \textbf{129}, 239902 (2022).
\bibitem{arXiv:2108.02525} F. Anzuini, N. F. Bell, G. Busoni, et al., J. Cosmol. Astropart. Phys. \textbf{11}, 056 (2021).
\bibitem{arXiv:2203.02132} A. Kumar, H. C. Das, and S. K. Patra, Mon. Not. R. Astron. Soc. \textbf{513}, 1820 (2022).
\bibitem{arXiv:2204.02238} M. Fujiwara, K. Hamaguchi, N. Nagata, and J. Zheng, Phys. Rev. D \textbf{106}, 055031 (2022).
\bibitem{arXiv:2104.02700} T. N. Maity and F. S. Queiroz, Phys. Rev. D \textbf{104}, 083019 (2021).
\bibitem{arXiv:1807.02840} N. F. Bell, G. Busoni, and S. Robles, J. Cosmol. Astropart. Phys. \textbf{09}, 018 (2018).
\bibitem{arXiv:2301.08767} G. Alvarez, A. Joglekar, M. Phoroutan-Mehr, and H. -B. Yu, Phys. Rev. D \textbf{107}, 103024 (2023).
\bibitem{arXiv:1904.09803} N. F. Bell, G. Busoni, and S. Robles, J. Cosmol. Astropart. Phys. \textbf{06}, 054 (2019).
\bibitem{arXiv:1906.10145} R. Garani and J. Heeck, Phys. Rev. D \textbf{100}, 035039 (2019).
\bibitem{arXiv:1911.06334} J. F. Acevedo, J. Bramante, R. K. Leane, and N. Raj, J. Cosmol. Astropart. Phys. \textbf{03}, 038 (2020).
\bibitem{arXiv:2001.09140} W. -Y. Keung, D. Marfatia, and P. -Y. Tseng, JHEP \textbf{07}, 181 (2020).
\bibitem{arXiv:2207.02221} J. Coffey, D. McKeen, D. E. Morrissey, and N. Raj, Phys. Rev. D \textbf{106}, 115019 (2022).
\bibitem{arXiv:2002.10961} K. Zhang, G. -Z. Huang, J. -S. Tsao, and F. -L. Lin, Eur. Phys. J. C \textbf{82}, 366 (2022).
\bibitem{arXiv:2109.01853} H. C. Das, A. Kumar, and S. K. Patra, Phys. Rev. D \textbf{104}, 063028 (2021).
\bibitem{arXiv:2110.05538} B. K. K. Lee, M. -c. Chu, and L. -M. Lin, Astrophys. J. \textbf{922}, 242 (2021).
\bibitem{arXiv:1905.00893} A. Herrero, M. A. Perez-Garcia, J. Silk, and C. Albertus, Phys. Rev. D \textbf{100}, 103019 (2019).
\bibitem{arXiv:1905.12483} S. A. Bhat and A. Paul, Eur. Phys. J. C \textbf{80}, 544 (2020).
\bibitem{arXiv:2002.00594} H. C. Das, A. Kumar, B. Kumar, et al., Mon. Not. R. Astron. Soc. \textbf{495}, 4893 (2020).
\bibitem{arXiv:2212.12615} C. H. Lenzi, M. Dutra, O. Lourenco, et al., Eur. Phys. J. C \textbf{83}, 266 (2023).
\bibitem{arXiv:2209.10905} E. Giangrandi, V. Sagun, O. Ivanytskyi, C. Providencia, and T. Dietrich, Astrophys. J. \textbf{953}, 115 (2023).
\bibitem{arXiv:2304.05100} P. Routaray, S. R. Mohanty, H. C. Das, et al., J. Cosmol. Astropart. Phys. \textbf{10}, 073 (2023).
\bibitem{arXiv:2306.17510} V. Parmar, H. C. Das, M. K. Sharma, and S. K. Patra, Phys. Rev. D \textbf{108}, 083003 (2023).
\bibitem{arXiv:2311.00113} M. Deliyergiyev, A. D. Popolo, and M. L. Delliou, Mon. Not. R. Astron. Soc. \textbf{527}, 4483 (2024).
\bibitem{arXiv:2007.05382} H. C. Das, A. Kumar, B. Kumar, S. K. Biswal, and S. K. Patra, J. Cosmol. Astropart. Phys. \textbf{01}, 007 (2021).
\bibitem{arXiv:2109.03801} D. Rafiei Karkevandi, S. Shakeri, V. Sagun, and O. Ivanytskyi, Phys. Rev. D \textbf{105}, 023001 (2022).
\bibitem{arXiv:2204.05560} Z. Miao, Y. Zhu, A. Li, and F. Huang, Astrophys. J. \textbf{936}, 69 (2022).
\bibitem{arXiv:1910.09925} O. Ivanytskyi, V. Sagun, and I. Lopes, Phys. Rev. D \textbf{102}, 063028 (2020).
\bibitem{arXiv:2212.12547} T. T. Q. Nguyen and T. M. P. Tait, Phys. Rev. D \textbf{107}, 115016 (2023).
\bibitem{arXiv:1806.00568} H. Sotani and K. D. Kokkotas, Phys. Rev. D \textbf{97}, 124034 (2018).
\bibitem{arXiv:1808.04391} H. O. Silva and N. Yunes, Phys. Rev. D \textbf{99}, 044034 (2019).
\bibitem{arXiv:2007.10080} R. Xu, Y. Gao, and L. Shao, Phys. Rev. D \textbf{102}, 064057 (2020).
\bibitem{arXiv:2109.13453} Z. Hu, Y. Gao, R. Xu, and L. Shao, Phys. Rev. D \textbf{104}, 104014 (2021).
\bibitem{arXiv:2005.12758} J. Soldateschi, N. Bucciantini, and L. Del Zanna, Astron. Astrophys. \textbf{640}, A44 (2020).
\bibitem{arXiv:2010.14833} J. Soldateschi, N. Bucciantini, and L. Del Zanna, Astron. Astrophys. \textbf{645}, A39 (2021).
\bibitem{arXiv:2004.00322} N. Bucciantini and J. Soldateschi, Mon. Not. R. Astron. Soc. \textbf{495}, L56 (2020).
\bibitem{arXiv:2107.07036} R. F. P. Mendes, N. Ortiz, and N. Stergioulas, Phys. Rev. D \textbf{104}, 104036 (2021).
\bibitem{arXiv:2105.13644} R. Niu, X. Zhang, B. Wang, and W. Zhao, Astrophys. J. \textbf{921}, 149 (2021).
\bibitem{arXiv:1903.00391} A. S. Arapoglu, K. Y. Eksi, and A. E. Yukselci, Phys. Rev. D \textbf{99}, 064055 (2019).
\bibitem{arXiv:2204.02138} S. Tuna, K. I. Unluturk, and F. M. Ramazanoglu, Phys. Rev. D \textbf{105}, 124070 (2022).
\bibitem{arXiv:1707.02809} S. Morisaki and T. Suyama, Phys. Rev. D \textbf{96}, 084026 (2017).
\bibitem{arXiv:1805.07818} K. V. Staykov, D. Popchev, D. D. Doneva, and S. S. Yazadjiev, Eur. Phys. J. C \textbf{78}, 586 (2018).
\bibitem{arXiv:1812.00347} D. Popchev, K. V. Staykov, D. D. Doneva, and S. S. Yazadjiev, Eur. Phys. J. C \textbf{79}, 178 (2019).
\bibitem{arXiv:1712.03715} D. D. Doneva and S. S. Yazadjiev, J. Cosmol. Astropart. Phys. \textbf{04}, 011 (2018).
\bibitem{arXiv:2103.11999}  H. -J. Kuan, D. D. Doneva, and S. S. Yazadjiev, Phys. Rev. Lett. \textbf{127}, 161103 (2021).
\bibitem{arXiv:1911.06908} D. D. Doneva and S. S. Yazadjiev, Phys. Rev. D \textbf{101}, 064072 (2020).
\bibitem{arXiv:2004.03956} D. D. Doneva and S. S. Yazadjiev, Phys. Rev. D \textbf{101}, 104010 (2020).
\bibitem{arXiv:2105.08543} H. -J. Kuan, J. Singh, D. D. Doneva, et al., Phys. Rev. D \textbf{104}, 124013 (2021).
\bibitem{arXiv:2109.09399} K. V. Staykov and R. Z. Zheleva, Eur. Phys. J. C \textbf{82}, 108 (2022).
\bibitem{arXiv:2207.13624} H. Boumaza and D. Langlois, Phys. Rev. D \textbf{106}, 084053 (2022).
\bibitem{Barranco} J. Barranco, A. Bernal, and D. Nunez, Mon. Not. R. Astron. Soc. \textbf{449}, 403 (2015).
\bibitem{Rezaei} Z. Rezaei, Astrophys. J. \textbf{835}, 33 (2017).
\bibitem{Narain} G. Narain, J. Schaffner-Bielich, and I. N. Mishustin, Phys. Rev. D \textbf{74}, 063003 (2006).
\bibitem{28} R.F.P. Mendes and N. Ortiz, Phys. Rev. D \textbf{93}, 124035 (2016).
\bibitem{Jiang} J. -L. Jiang, et al., Astrophys. J. \textbf{885}, 39 (2019).
\bibitem{Ozelprd} F. Ozel and D. Psaltis, Phys. Rev. D \textbf{80}, 103003 (2009).
\bibitem{26} T. Damour and G. Esposito-Farese, Phys. Rev. Lett. \textbf{70}, 2220 (1993).
\bibitem{Ozel} F. Ozel, T. Guver, and D. Psaltis, Astrophys. J. \textbf{693}, 1775 (2009).
\bibitem{Guver} T. Guver, P. Wroblewski, L. Camarota, and F. Ozel, Astrophys. J. \textbf{719}, 1807 (2010).
\bibitem{Abbott7}  B. P. Abbott, et al., Phys. Rev. Lett. \textbf{119}, 161101 (2017).
\bibitem{Abbott8} B. P. Abbott, et al., Phys. Rev. Lett. \textbf{121}, 161101 (2018).
\bibitem{Miller} M. C. Miller, et al., Astrophys. J. lett. \textbf{887}, L24 (2019).

\end{thebibliography}
\end{document}